\documentclass[twocolumn,nofootinbib,showpacs,prd,superscriptaddress]{revtex4}
\usepackage{amsmath}
\usepackage{amsfonts}
\usepackage{amssymb}
\usepackage{graphicx}
\usepackage{color}

\begin{document}

\title{Cosmological degeneracy versus cosmography: a cosmographic dark energy model}

\author{Orlando Luongo}
\email{luongo@na.infn.it}
\affiliation{Department of Mathematics and Applied Mathematics, University of Cape Town, Rondebosch 7701, Cape Town, South Africa.}
\affiliation{Astrophysics, Cosmology and Gravity Centre (ACGC), University of Cape Town, Rondebosch 7701, Cape Town, South Africa.}
\affiliation{Dipartimento di Fisica, Universit\`a di Napoli ''Federico II'', Via Cinthia, I-80126, Napoli, Italy.}
\affiliation{Istituto Nazionale di Fisica Nucleare (INFN), Sez. di Napoli, Via Cinthia, Napoli, Italy.}

\author{Giovanni Battista Pisani}
\affiliation{Dipartimento  di Fisica, ``Sapienza" Universit\`a di Roma, Piazzale Aldo Moro, 5, I-00185, Roma, Italy.}

\author{Antonio Troisi}
\affiliation{Dipartimento di Fisica ``E.R. Caianiello", Universit\`a di Salerno, Via Giovanni Paolo II, 132, 84084, Salerno, Italy.}

\begin{abstract}
In this work we use cosmography to alleviate the degeneracy among cosmological models, proposing a way to parameterize matter and dark energy in terms of cosmokinematics quantities. The recipe of using cosmography allows to expand observable quantities in Taylor series and to directly compare those expansions with data. We adopt this strategy and we propose a fully self-consistent parametrization of the total energy density driving the late time universe speed up. Afterwards, we describe a feasible \emph{cosmographic dark energy model}, in which matter is fixed whereas dark energy evolves by means of the cosmographic series. Our technique provides robust constraints on cosmokinematic parameters, permitting one to separately bound matter from dark energy densities. Our cosmographic dark energy model turns out to be one parameter only, but differently from the $\Lambda$CDM paradigm, it does not contain ansatz on the dark energy form. In addition, we even determine the free parameter of our model in suitable $1\sigma$ intervals through Monte Carlo analyses based on the Metropolis algorithm. We compare our results with the standard concordance model and we find that our treatment seems to indicate that dark energy slightly evolves in time, reducing to a pure cosmological constant only as $z\rightarrow0$.
\end{abstract}

\pacs{98.80.-k, 98.80.Jk, 98.80.Es}

\maketitle


\section{Introduction}

Observations of type Ia supernovae (SNeIa) undoubtedly portrayed a late-time speeding up universe \cite{RiessPerlmutter,SNeIa-2,SNeIa-3}, dominated by some sort of exotic fluid, different from matter and responsible for the cosmic acceleration. This component, dubbed dark energy, exhibits a anti-gravitational negative equation of state (EoS) \cite{ga}. Enclosing the corresponding dark energy density within Einstein's energy momentum tensor means to counterbalance the gravitational attraction, reproducing cosmological observations. Several explanations have been carried out in the literature in order to motivate dark energy's existence, albeit a complete and self consistent physical interpretation is so far unknown \cite{burgess2013,peebles2003,deputter2010,dent2013,copeland}.

\noindent The simplest approach leads to introducing a cosmological constant, i.e. $\Lambda$, associated to quantum field vacuum energy. The corresponding paradigm, namely the $\Lambda$CDM model \cite{sahni1999} excellently fits cosmological data, providing $\Lambda$ to be the most promising dark energy explanation. Notwithstanding its experimental successes, the $\Lambda$CDM model suffers from a profound \emph{fine-tuning problem} \cite{weilambda}. Particularly, quantum field predictions suggests $\Lambda$ to differ from 120 orders of magnitude from cosmological measurements. In addition, comparable magnitudes between matter and $\Lambda$ densities have been experimentally found, leading to a serious \emph{coincidence problem} \cite{coi}. This coincidence may be healed if one supposes an evolving dark energy EoS, whereas fine-tuning is avoided if one does not assume $\Lambda$'s existence. For those reasons, a \emph{plethora} of alternative approaches have been proposed throughout the years, spanning from modifications of the Hilbert-Einstein action \cite{cliftonR2011}, additional scalar or tachyonic fields \cite{bagla2003}, K-essence approach \cite{ArmendarizPicon2000}, to parameterizations of the barotropic factor, phenomenological pressures, varying cosmological constant \cite{zlatev1998}, holographic principle \cite{miao2004} and so forth \cite{cho2012}.

\noindent Unfortunately, all dark energy models are plagued from the problem of separately measuring present-time matter density and total EoS, $\Omega_{m,0}$ and $\omega_0$ respectively. Rephrasing it, measurements of $\Omega_{m,0}$ are \emph{intertwined} with allowed values of $\omega_0$ or with curvatures in models with no definite ansatz on the Friedmann flow \cite{degenerazione1,degenerazione2,degenerazione3,rubano}. Thus, all energy densities enter the Hubble flow indistinguishably and so, a strong degeneracy occurs if one simultaneously measures different cosmological fluids \cite{ratra,linder2011,degenerazione4}. This causes that different models are capable of fitting data with high accuracy, whereby it is impossible to univocally determine the dark energy physical nature. This is due to the fact that, for vanishing spatial curvature, a homogeneous and isotropic universe is described by one function only, i.e. the scale factor $a(t)$, accounting  the \emph{total} universe energy budget. A {\it dark degeneracy} problem arises, representing a disturbing conundrum to establish properties of dark energy. The strategy of separate measurements of $\Omega_{m,0}$ and $\omega_0$ suffers from an interdependence that does not allow to draw distinct information on both matter and dark energy. The energy-momentum tensor is postulated to satisfy geometric prescriptions, related to Bianchi identities and, in the coarse-grained case of perfect fluids, it reads $\mathcal T_{\mu\nu}\,\equiv\,\mbox{diag}(\rho(t),\,-\mathcal P(t),\,-\mathcal P(t),\,-\mathcal P(t))$. As a technical consequence, Friedmann equations account for the  pressure and densities of all components involving in the universe description. Hence, we only measure the total energy density, instead of single components. The total EoS is simply defined as $\omega\equiv{\mathcal P \over \rho}$, sometimes referred to as the \emph{barotropic factor} if $\omega$ does not depend explicitly on the entropy.

In this paper, we first propose a cosmographic method for alleviating the dark degeneracy. Second, we present a \emph{cosmographic dark energy model}, which permits one to parameterize separately matter from dark energy in a scheme that depends only on cosmokinematic measurable quantities.
Our technique allows to fix robust constraints on observables, regardless of any priors imposed from the beginning on $\Omega_{m,0}$. We only need the cosmological principle to circumvent degeneracy by means of a fully kinematic dark energy reconstruction. The method is well supported by cosmographic demands, which represent a way to bound cosmological constraints involving the scale factor derivatives only, in a model independent framework \cite{visser2004,cattoen2008,bamba2012}. In fact, since scale factor derivatives can be accurately measured up to a certain order, cosmography fixes in turn $\omega_0$ and $\Omega_{m,0}$ employing only two cosmographic coefficients: the universe acceleration, $q(z)$, and the variation of acceleration, $j(z)$. Further, the assumption of a completely general expression for the Hubble rate $H(z)$ without considering any prescription on its dynamical behavior can be naturally derived. As a result, it is possible to reduce the number of degrees of freedom related to the degeneracy problem. In particular, since degeneracy arises as a byproduct of the standard lore dependence on a single function, our model is built up by means of cosmographic parameters only, without any conjecture on the dark energy evolution. In the last part of our work, in fact, we compare our approaches with cosmological data, getting limits over the cosmographic free parameters which are model-independent quantities. Thus, these procedures may alleviate the dark degeneracy and suggest a way to \emph{disentangle} the measurements of $\omega_0$ and $\Omega_{m,0}$. A high-impact application of our treatment deals with determining whether dark energy evolves or not in time. In our framework, we get a possible dark energy evolution which would be different from a pure cosmological constant.

The paper is organized as follows. Section II is dedicated to resume the cosmic degeneracy and to discussion on the basic requirements of cosmography. Section III is devoted to explain how to alleviate the degeneracy problem by means of cosmography. In the same section, we propose the definition of our cosmographic model. Section IV deals with data fitting procedures performed to deduce experimental constraints. Section V finally regards comments and conclusions with particular attention to the perspectives of our approach.


\section{Cosmography versus Degeneracy}\label{secondpar}

Here, we present how cosmography can alleviate the cosmic degeneracy between the total EoS and matter density. To do so, we need to enter the technical aspects of  the degeneracy issue, showing the strategy to relate it to cosmography. First, we circumscribe our analysis to a spatially flat universe ($k\,=\,0$), employing the Friedmann-Robertson-Walker (FRW) metric, i.e.
$\displaystyle{ds^2=dt^2-a(t)^2\left(dr^2+r^2d\Omega\right)}$, with $d\Omega\equiv d\theta^2+\sin^2{\theta}d\phi^2$. Thus, the Friedmann equations read\,:
\begin{equation}
\label{Fried}
H^2 = \left(\frac{\dot{a}}{a}\right)^2  = \frac{8\pi G}{3}\rho\,,
\end{equation}
\begin{equation}\label{Fried2}
\dot{H}+H^2 = \frac{\ddot{a}}{a} = -\frac{4\pi
G}{3}\left(\rho+3\mathcal P\right)\,,
\end{equation}
where $H=\frac{d}{dt}\ln a(t)$ is the Hubble parameter. Any cosmological model is univocally characterized once $\mathcal P=\mathcal P(\rho)$ is determined. Unfortunately, feasible bounds on the \emph{total} EoS are unable to experimentally split the barotropic factor into components. Saying it differently, since $\sum_i\omega_i\neq \omega$, with the subscript $i$ indicating each cosmic species, it seems difficult to impose limits either on $\rho_i$ or $\omega_i$ by considering numerical bounds on $\omega$ only.

Generally, at present time, there is a common arbitrariness in splitting $\omega$ into (at least) two species: the first is clearly due to dark energy, whereas the other one to the universe total matter. Any further contributions, e.g. neutrinos, photons, string relics, etc. do not significantly contribute to the whole energy budget at $z\simeq0$ and may be neglected. Combining between them the Friedmann equations and introducing the density parameter $\Omega_{m,0}\,=\displaystyle\,\frac{\rho_m}{\rho_{crit}}$ for matter, one obtains a general expression for the EoS of dark energy \cite{rubano}:
\begin{equation}\label{eos}
\omega=-\frac{1}{3}\left(\frac{2\dot
H+3H^2}{H^2-H_{0}^{2}\Omega_{m,0}a^{-3}}\right)\,.
\end{equation}
Equation \eqref{eos} suggests that, given a range of intervals for $\Omega_{m,0}$, the term $\omega$ spans a corresponding tight interval of allowed values. Even tracing the universe expansion history does not enable to separate both those intervals, therefore it is hard to discriminate the best one among different cosmological models  \cite{kunz}. In other words, several cosmological paradigms constitute equivalent classes of models capable of describing the universe dynamics at late-times.

\noindent We want to show here that, through the use of cosmography, one can \emph{circumvent} the measurement of $\Omega_{m,0}$ and $\omega$ separately, without fixing the cosmological model \emph{a priori} \cite{degenerazioneultima}. To do so, let us introduce the basic requirements of cosmography. In particular, cosmography represents a branch of cosmology, which only involves the use of easygoing symmetry assumptions, concerning the validity of the cosmological principle. A feasible consequence of this prescription is expanding the scale factor $a(t)$ into power series around $t=t_0$. So that, we obtain

\begin{eqnarray}\label{serie1}
a(t)= 1 + H_0 \Delta t - {1\over2} q_0 H_0^2 \Delta t ^2 +{1\over6} j_0 H_0^3\Delta t ^3 + \ldots\,,
\end{eqnarray}
with $\Delta t\,=\,t-t_0$. The strategy behind the above expansion provides the relevant property of cosmography of being model independent in choosing bounds on derivatives of $a(t)$ \cite{visser2004,posy,luongoprd2012}. To better understand this fact, the scale factor derivatives may be set as follows
\begin{equation}\label{q_j_definition}
\left\{
\begin{array}{lll}
q = -\frac{1}{H^2} \frac{\ddot{a}}{a}\ & & q(t)=-\frac{\dot{H}}{H^2} -1\,  \\
& \Leftrightarrow & \,\\
j = \frac{1}{H^3} \frac{a^{(3)}}{a}\ & & j(t)=\frac{\ddot{H}}{H^3}-3q-2\,,
\end{array}
\right.
\end{equation}
which have been also expressed in terms of the measurable quantity $H$. These quantities are known in the literature as the acceleration parameter, $q$, quantifying whether the universe accelerates or decelerates,  and the jerk parameter, $j$, showing whether the universe changes its acceleration sign after the transition time \cite{transition1,transition2}.
In our picture, the subscript $0$ in Eqs. ($\ref{serie1}$) underlines that $H,q$ and $j$ are presently fixed, i.e. $z=0$. Hence, one argues cosmography can fix important constraints on our present universe. The reasons which permit to conclude that cosmography does not depend on the choice of any cosmological model are listed as follows \cite{reviewmia}.
\begin{itemize}
  \item Scalar curvature is somehow fixed to be zero. Precise observations at late and early-times confirm this property, showing that the universe is spatially flat. The measurement of the jerk parameter depends on the spatial curvature $\Omega_k$, but if $\Omega_k$ vanishes, the jerk parameter is purely model-independently determined.
  \item All observable quantities can be expressed in terms of the scale factor or alternatively by means of the redshift $z$, through the formula $a=\frac{1}{1+z}$. This fact is a technical advantage of cosmography, since different data sets well adapt to cosmographic analyses.
\end{itemize}
Further, we commonly refer to the \emph{cosmographic series} (CS) as the set of numerical constraints on $H,q$ and $j$ evaluated today. The need of precise bounds on CS becomes essential to extrapolate information on the universe dynamics. Indeed, it has been shown that Taylor expansions are plagued by a severe convergence problem, since expansions are based on assuming that $z\sim0$, whereas cosmological data typically exceed this range. Hence, without deeply entering this argument, to reduce any systematics in the fitting procedures, we develop all series expansions up to a suitable order (see for details \cite{luongoprd2012}). For those reasons, we only consider the first three cosmographic terms in Eq. (\ref{serie1}), i.e. $H, q, j$, since they are more efficiently constrained by observational data.

\noindent We now consider a sort of ``back scattering" procedure, to fix the free parameters of a dark energy model, without the need of finding out constraints on $\Omega_{m,0}$ and $\omega_0$. In order to build up our model, we rewrite $\Omega_{m,0}$ and $\omega_0$ in terms of cosmographic quantities. We therefore obtain a sort of ``fully" cosmographic Hubble flow \cite{cattoen2008} which is not plagued by the initial bias due to the particular choice of dark energy density. The procedure to obtain a fully cosmographic dark energy model consists in possible parameterizations of the EoS of dark energy through the use of the CS. This treatment substitutes the measurement of $\Omega_{m,0}$ with the observable value of $q$ and $j$ at present time. As we will clarify in the next section, since $q_0$ and $j_0$ may be bounded independently, by simply comparing the expanded luminosity distance directly with data, one may get a Hubble rate parametrization in which the matter contribute is replaced by cosmographic quantities. We also quantitatively demonstrate that this method does not shift the degeneracy from matter and $\omega$ to the CS, but allows to fix limits on a new set of cosmographic variables which are bounded directly with data.


\section{The \emph{cosmographic model}: how to split the EoS from cosmography}

In this section, we focus on how to build up our \emph{cosmographic model}. The only basic requirement consists in splitting the matter fluid from the dark energy term in the Hubble rate, saying that:
\begin{equation}\label{hz}
H(z)=H_0\sqrt{\Omega_{m,0}(1+z)^{3}+\Omega_{DE}G(z)}\,.
\end{equation}
In Eq. (\ref{hz}), the dark energy density $\Omega_{DE}$ is unknown and it is supposed to vary in terms of the redshift $z$ through a non specified function $G(z)$, built up by the following properties:

\begin{equation}
\left\{
                 \begin{array}{ll}
                   G(z)\rightarrow 1 \,,\quad & \hbox{$z=0$;} \\
                   \,\\
                   \Omega_{DE}\equiv1-\Omega_{m,0}\,,\quad & \hbox{$\forall z$;} \\
                   \,\\
                   \displaystyle\frac{G(z)}{(1+z)^3} \gtrsim\frac{\Omega_{m,0}}{\Omega_{DE}}\,,\quad & \hbox{$z\rightarrow 0$.}
                 \end{array}
               \right.
\end{equation}
The first condition, reported in the above list, leads to $H=H_0$ as $z=0$. The second condition is a direct consequence of the first condition, while the third condition requires that dark energy dominates over matter at late-times. Those three properties derive from the splitting of the net energy-momentum tensor and are compatible with the standard cosmological requirements.

Let us consider  Eq. ($\ref{hz}$) and the definitions of $q$
and $j$ in terms of $H$, given by Eqs. ($\ref{q_j_definition}$). Merging these two relations, one obtains
\begin{widetext}
\begin{equation}\label{qudef}
q(z)=-1+\frac{(1+z)\left[3\Omega_{m,0}(1+z)^{2}+\Omega_{DE}G'(z)\right]}{2\left[\Omega_{m,0}(1+z)^{3}+\Omega_{DE}G(z)\right]}\,,
\end{equation}
and
\begin{equation}\label{jeidef}
j(z)=\frac{2\Omega_{DE}G(z)+(1+z)\left\{-2\Omega_{DE}G'(z)+(1+z)\left[2\Omega_{m,0}(1+z)+\Omega_{DE}G''(z)\right]\right\}}{2\left[\Omega_{m,0}(1+z)^{3}+\Omega_{DE}G(z)\right]}\,,
\end{equation}
\end{widetext}
which describe the cosmographic parameters in terms of matter and dark energy contents of the universe. Here, apices $'$ represent derivatives with respect to the redshift $z$. In practice, cosmography gives the possibility to  replace the mass density through the values of $q_0$, $j_0$  exploiting Eqs. (\ref{qudef}) and (\ref{jeidef}).

\noindent Our second step is to expand $q$ and $j$, around $a=1$:
\begin{subequations}
\begin{align}
\label{expans_q_a}
q\,=\,\sum_{k=0}^{\infty}\frac{1}{k!}\frac{d^kq}{da^k}(1-a)^k\,,\\
j\,=\,\sum_{k=0}^{\infty}\frac{1}{k!}\frac{d^kj}{da^k}(1-a)^k\,,
\end{align}
\end{subequations}
which clearly reduce to $q=q_0$ and $j=j_0$, as $a=1$. 
For our purposes, it behooves us to truncate the above series at the second order in $a$. We therefore get:
\begin{subequations}
\begin{align}
q(z)\,=\,q_0 + q_1\frac{z}{1+z} + q_2\left(\frac{z}{1+z}\right)^2+\ldots\,,\label{expans_q_z}\\
j(z)\,=\,j_0+j_1\frac{z}{1+z} + j_2\left(\frac{z}{1+z}\right)^2+\ldots\,.\label{expans_j_z}
\end{align}
\end{subequations}
It is easy to notice that the signs of $q_1,q_2,j_1,j_2\ldots$ are not directly fixed from the expansion. However, from theoretical arguments we can find out allowed intervals for those terms. For example, if we assume that the above expansions are valid at different stages of the universe evolution, we conclude that $q(z)$ reduces to $q\rightarrow q_0+q_1+q_2>0$ for $z\rightarrow\infty$. On the other side, to account for the dark energy effects at late times, one expects $-1<q_0<0$, while $j_0$ is expected to be positive in order to guarantee that $q$ could change its sign in time \cite{posy}. To preserve this behavior, it is strictly necessary to assume that $j_0>0$ thus we will adopt this assumption in the following. Thence, by using Eq. ($\ref{qudef}$) and the series expansions Eqs. (\ref{expans_q_z}) - (\ref{expans_j_z}), we infer

\begin{eqnarray}\label{darkevolution}
G(z)=\frac{1}{\Omega_{DE}}\Big[(1+z)^\alpha e^{-\beta(z)}-\Omega_{m,0}(1+z)^{3}\Big]\,,
\end{eqnarray}
with $\alpha={2(1+q_{0}+q_{1}+q_{2})}$ and
\begin{equation}
\beta(z)=\frac{z\left[2 q_{1} (1 + z) + q_{2} (2 + 3 z)\right]}{(1 + z)^2}.
\end{equation}
Relation ($\ref{darkevolution}$) shows the interesting property that the Hubble rate $H(z)$ is independent from $\Omega_m(z)$ at every redshift, in fact the general cosmographic expression for $H(z)$ reads:

\begin{equation}\label{extendedH}
H\,=\,H_0\,\left[\frac{(1+z)^{\alpha}}{e^{\beta(z)}}\right]^{\frac{1}{2}}\,,
\end{equation}

\noindent providing the advantage that our model turns out to be mass-independent. Analogously, even the effective barotropic factor $\omega(z)$ becomes mass-independent, as it will be clarified later in the text.

\noindent In principle, our approach is a three-parameter model. However, combining Eqs. ($\ref{jeidef}$) and ($\ref{darkevolution}$), we have
\begin{equation}\label{jzz}
j(z)=2q(z)^2+q(z)+\frac{q_1}{1+z}+2q_2\frac{z}{(1+z)^2}\,,
\end{equation}
which fixes constraints, together with Eqs. \eqref{expans_q_z} - \eqref{expans_j_z}, among $j_0$, $j_1$, $j_2$ and $q_0$, $q_1$ and $q_2$, giving:
\begin{subequations} \label{tutto}
\begin{align}
j_{0}\,&=\,q_{0}+2q_{0}^2+q_{1}\,,\label{relazione1}\\
j_{1}\,&=\,4q_{0}q_{1}+2q_{2}\,,\label{relazione2}\\
j_{2}\,&=\,2q_{1}^2+4q_{0}q_{2}-q_{2}\label{relazione3}\,.
\end{align}
\end{subequations}
To get limits over $q_0$ and $j_0$ in a model-independent way, one can simply consider the standard luminosity distance in function of the redshift as:
\begin{equation}\label{stp}
 d_L  = \frac{1}{a(t)}\int_{t}^{t_0}{\frac{dt'}{a(t')}} \,=\,(1+z)\int_{0}^{z}\frac{dz'}{H(z')}\,,
\end{equation}
and so, rewriting it in terms of third order cosmographic parameters as\footnote{See \cite{luongoprd2012} for more general expansions.}, we have:
\begin{subequations}
\begin{align}
d_{L}^{(3)}  &\approx   \frac{z}{H_0} \left( 1 + \alpha_1\,z + \alpha_2\,z^2 + \ldots\right)\,,\label{lumdis1}\\
\alpha_1&=\frac{1}{2} - \frac{q_0}{2}\,,\label{lumdis2}\\
\alpha_2&=\frac{q_0^2}{2}-\frac{1}{6}  + \frac{q_0}{6} -\frac{j_0}{6}\,.\label{lumdis3}
\end{align}
\end{subequations}
It is evident that a direct comparison of $d_{L}^{(3)}$ with cosmological data permits to fix $q_0$ and $j_0$ in a model independent way. In fact, in Eq. (\ref{stp}), no information on the form of $H(z)$ have been introduced before expanding to get Eq. (\ref{lumdis1}). Clearly, observational measurements of $j_0$ and $q_0$ constrain the corresponding cosmographic parameters $q_1, j_1$. Suitable limits over $q_0,j_0$ have been reported in Tab. I, determining $q_1$ in a suitable confidence interval. The priors imposed over $q_0$ and $j_0$ have been postulated by means of recent developments on cosmography, e.g. for example \cite{ga,cattoen2008,posy,cattoencqg2007,luongo10}. In this paper, we are not interested in the standard procedure of fitting $q_0, j_0$ from Eq. \eqref{lumdis1}, but we only need the $q_0$ and $j_0$ priors to fix limits over $q_1$ and $j_1$. From Eq. (\ref{relazione1}), we can note that $q_1$ is univocally determined if one knows $q_0$ and $j_0$, for example by fitting Eq. (\ref{lumdis1}) with data. The proposed cosmographic model is therefore a \emph{one parameter only}, i.e. $q_2$. The difference with the $\Lambda$CDM case is that in our picture \emph{there is no assumption on the DE origin, but we are interested in reconstructing its possible evolution in time}. An important way to check the goodness of our model is to relate Eqs. (\ref{tutto}) to $\omega_0$ and its derivatives. In doing so, one invokes a direct correspondence between the universe EoS and the cosmographic coefficients, in order to fully get a single-parameter cosmological model, in which degeneracy is effectively alleviated. Hence, let us rewrite the barotropic factor $\omega(z)$ combining the Friedmann equations ($\ref{Fried}$)-($\ref{Fried2}$) and the Hubble flow (\ref{extendedH})
\begin{widetext}
\begin{equation}\label{wz}
\omega(z)=\frac{-1+2q_0(1+z)^{2}+z\left[2(q_1-1)+z(2q_1+2q_2-1)\right]}{3(1+z)^{2}}\,.
\end{equation}
\end{widetext}
Considering the first order Taylor expansion\footnote{We can suitably consider $z$ or $a$ as expansion variables at late times, since cosmological observables have the same converging rate in this regime.} of Eq. (\ref{wz}) and comparing it with $\omega(z)\sim \omega_0+\omega_1z+\ldots$, where $\omega_1\equiv\frac{d\omega}{dz}\Big|_{0}$, we find:
\begin{equation}\label{gequat}
\omega_1=\frac{2}{3}q_1\,,
\end{equation}
and then, we re-obtain $q_1\approx j_0-2q_0^2-q_0$ \cite{bravetti} which is equivalent to Eq. (\ref{relazione1}) and provides the correspondence between $\omega_1$ and the cosmographic coefficient $q_1$ by fact confirming our single-parameter model.

\noindent Considering the EoS expression (\ref{wz}) at current time, one obtains as effective barothropic factor:
\begin{eqnarray}\label{omegagraph}
\omega_0=-\frac{1}{3}(1-2q_0)\,,
\end{eqnarray}
which does not depend on $q_1$ and $q_2$ and can be properly evaluated, by knowing a prior on $q_0$. Summing up, the universe EoS and its first derivative $\omega_1$ are intimately related to $q_0$ and $j_0$. The mass term is removed in Eq. (\ref{extendedH}) and the single free coefficient is $q_2$. The barotropic equation for the pressure is $\mathcal P_0=\frac{1}{3}(-1+2q_0)\rho_{cr,0}$ and the dark energy evolution is framed by the coefficient $q_2$.

Another conceptual advantage of our approach is mainly based on how to experimentally compare the model with data. Indeed, as one fits a generic cosmological model, the hidden assumption is that the model under exam is (statistically) the \emph{best one}. In other words, choosing the dark energy evolution means to fix the likelihood function, because one supposes from the beginning that the corresponding universe expansion history is fixed according to $G(z)$. If one fixes the likelihood function by considering Eq. ($\ref{extendedH}$), the conceptual problem of fixing the dark energy evolution is removed, since one measures a cosmographic parameter $q_2$. \emph{This is a consequence of considering the dark energy evolution in terms of cosmographic model independent quantities.} In the next section, we summarize the fitting procedure adopted in this work outlining the results that we obtained. Finally, we discuss  the advantages and disadvantages of our approach.

\begin{center}
\begin{table}[ht]
\begin{small}
\caption{{\small \emph{Outline of priors adopted for the cosmographic
fits based on the union 2.1 compilation.}}}
\begin{tabular}{ccc}
\hline\hline
 $\qquad${\bf Parameter}$\qquad$    &  $\qquad$$\qquad$  &
$\qquad${\bf Priors}$\qquad$   \\
 \hline

 $H_0$  &  & $\{65,\ 76\}$ \\[1ex]

 $q_0$  &  & $\{-1.10,\ -0.06\}$ \\[1ex]

 $q_1$  &  & $\{0.5,\ 3\}$ \\[1ex]

 $q_2$  &  & $\{0.5,\ 3\}$ \\[1ex]

\hline\hline
\end{tabular}
\end{small}
\label{tabellao}
\end{table}
\end{center}


\begin{figure*}
\includegraphics[width=7in]{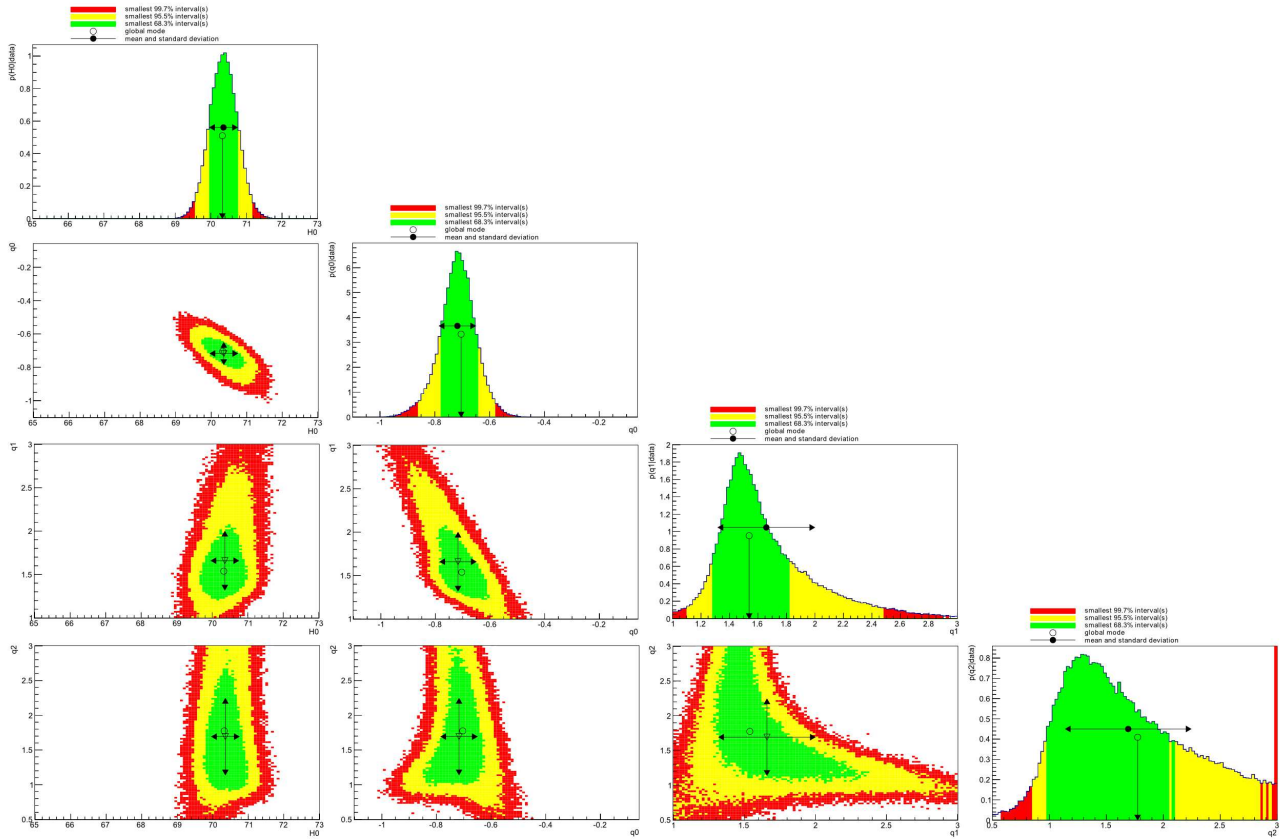}
\caption{\emph{Contour plots and likelihood functions obtained by directly fitting our theoretical distance modulus in terms of $q_0$,$q_1$ and $q_2$, as reported in Tab. \ref{tabellaq1q2}. A broadening in systematics is expected, although experimental results remain compatible with our second fit. }}
\label{ours}
\end{figure*}


\begin{figure*}[htb]
\includegraphics[width=6in]{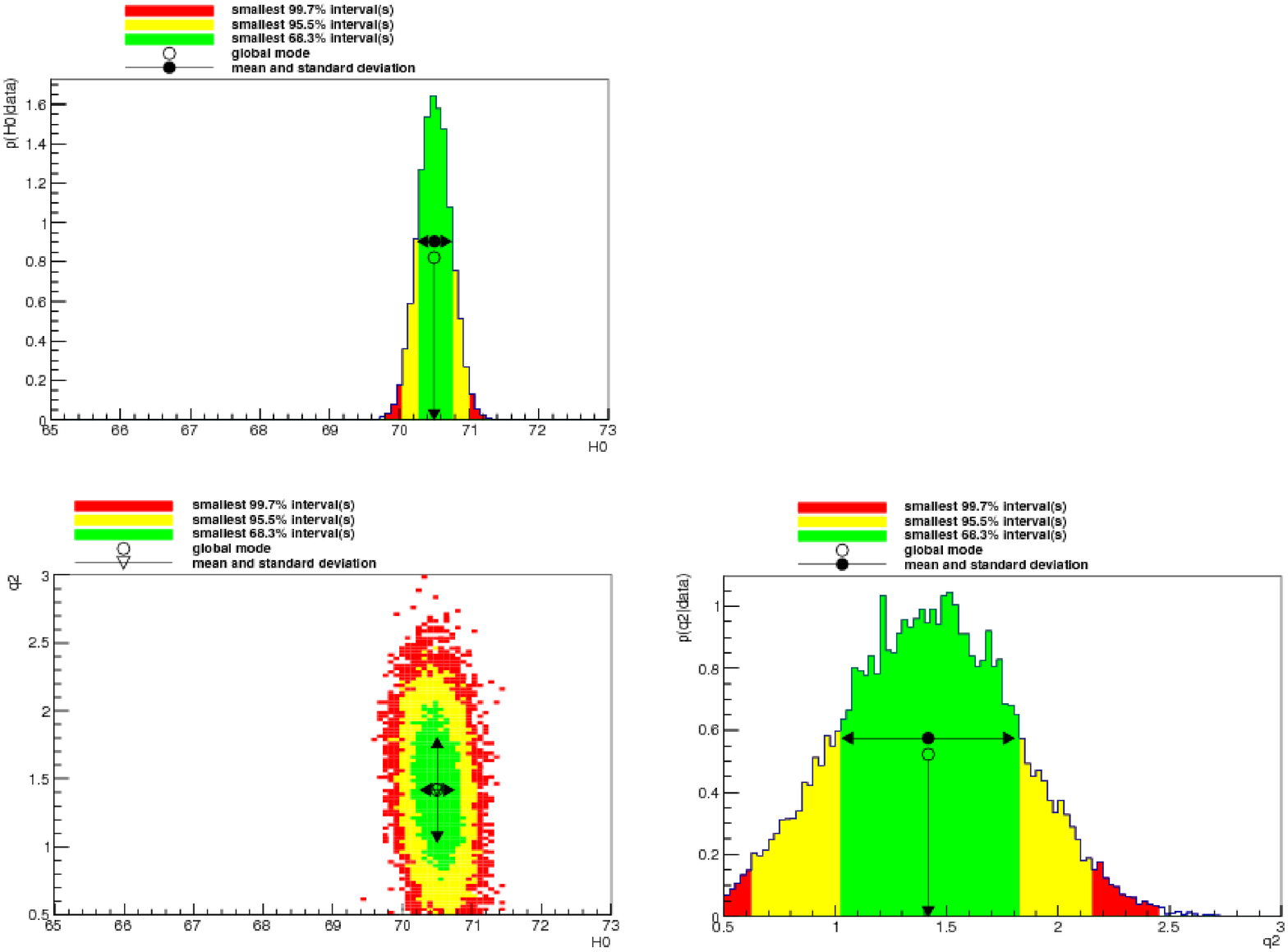}
\caption{\emph{Contours of the experimental analysis performed assuming $q_0$ and $q_1$ fixed with the values reported in Tab. \ref{tabellaDUE}. The numerical outcomes on $q_2$ are compatible with the ones of our first fit.}}
\label{ours2}
\end{figure*}


\section{The cosmographic fits}

We here present our strategies to obtain bounds on the observable quantities of the cosmographic model presented above. To show that, we recall the relevant fact that cosmography provides the advantage to constrain $H_0, q_0$ and $j_0$ alone, without invoking the form of a cosmological model a priori. This advantage turns out to give priors on $q_1, j_1$ and $j_2$ by simply solving Eqs. \eqref{tutto}. The only coefficient which remains unbounded is therefore $q_2$. For the sake of clearness, even the inverse procedure is allowed. In fact if one considers $q_0, q_1$ and $q_2$ in function of $j_0, j_1$ and $j_2$ the only free parameter could be $j_2$. For our purposes, it is much easier to perform numerical analyses treating $q_0$ and $j_0$ as model independent quantities and $q_1, q_2$ as free parameters of the model, with the possibility to get priors on $q_1$. This is due to the fact that employing $j_1, j_2$ as free parameters our treatment becomes more complicated and the corresponding equations to get $G(z)$ are not exactly solvable.

\noindent To determine cosmographic limits, we employ the use of the union 2.1 SNeIa compilation. We aim to perform two separate fits, in which we first fit the luminosity distance expanded around $z=0$ and we directly fit the quantities $q_0, q_1$ and $q_2$, whereas afterwards we fit $q_2$ taking $q_1$ fixed through the best fit of $q_0$ and $j_0$. The viable priors over coefficients have been involved into analysis by considering the upper and lower bounds on $q_0$ and $j_0$ and then calculating the other coefficients $q_1, q_2, j_1, j_2$ by means of Eqs. \eqref{lumdis1}. The spatial curvature is assumed to be negligibly small at late times as geometrical indications point out \cite{luongoprd2012} and we adopt the luminosity distance introduced in Eq. \eqref{lumdis1}. Although we do not exclude that our approach may be framed also in case of alternative distance definitions \cite{cattoencqg2007}, we do not consider alternative distances to constrain our cosmographic coefficients.

\begin{figure*}[htb]%
\centering
\parbox{3in}{%
\includegraphics[width=3in]{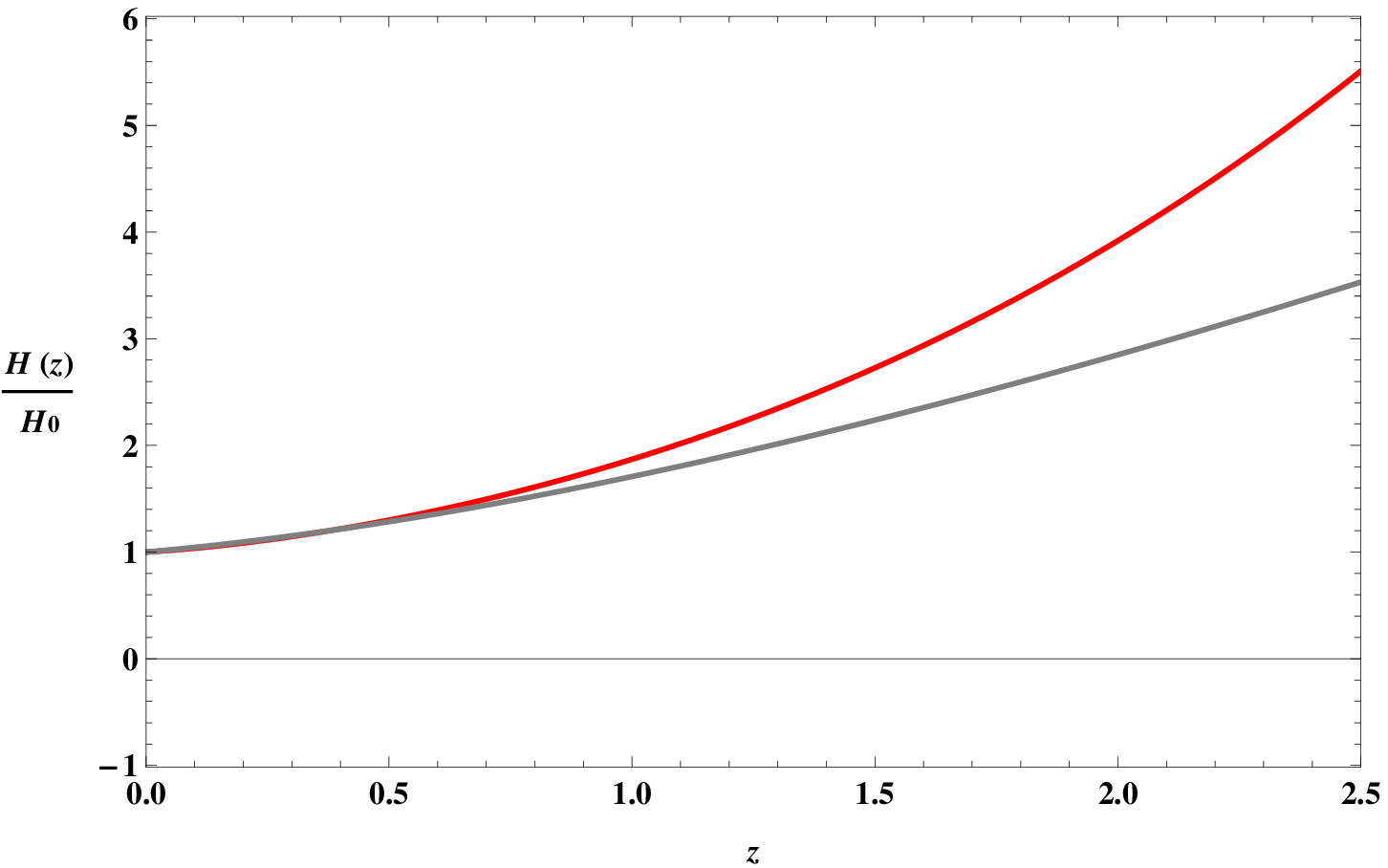}
\caption{\emph{Plot of the Hubble parameter in terms of the redshift $z$ considering the best fit values for $q_1$ and $q_2$ from the first fit. Solid red line refers to our model, while gray line is associated to the $\Lambda$CDM paradigm, with $\Omega_{m,0}=0.274$. \\}}%
\label{H}}%
\qquad\qquad
\parbox{3in}{%
\includegraphics[width=3in]{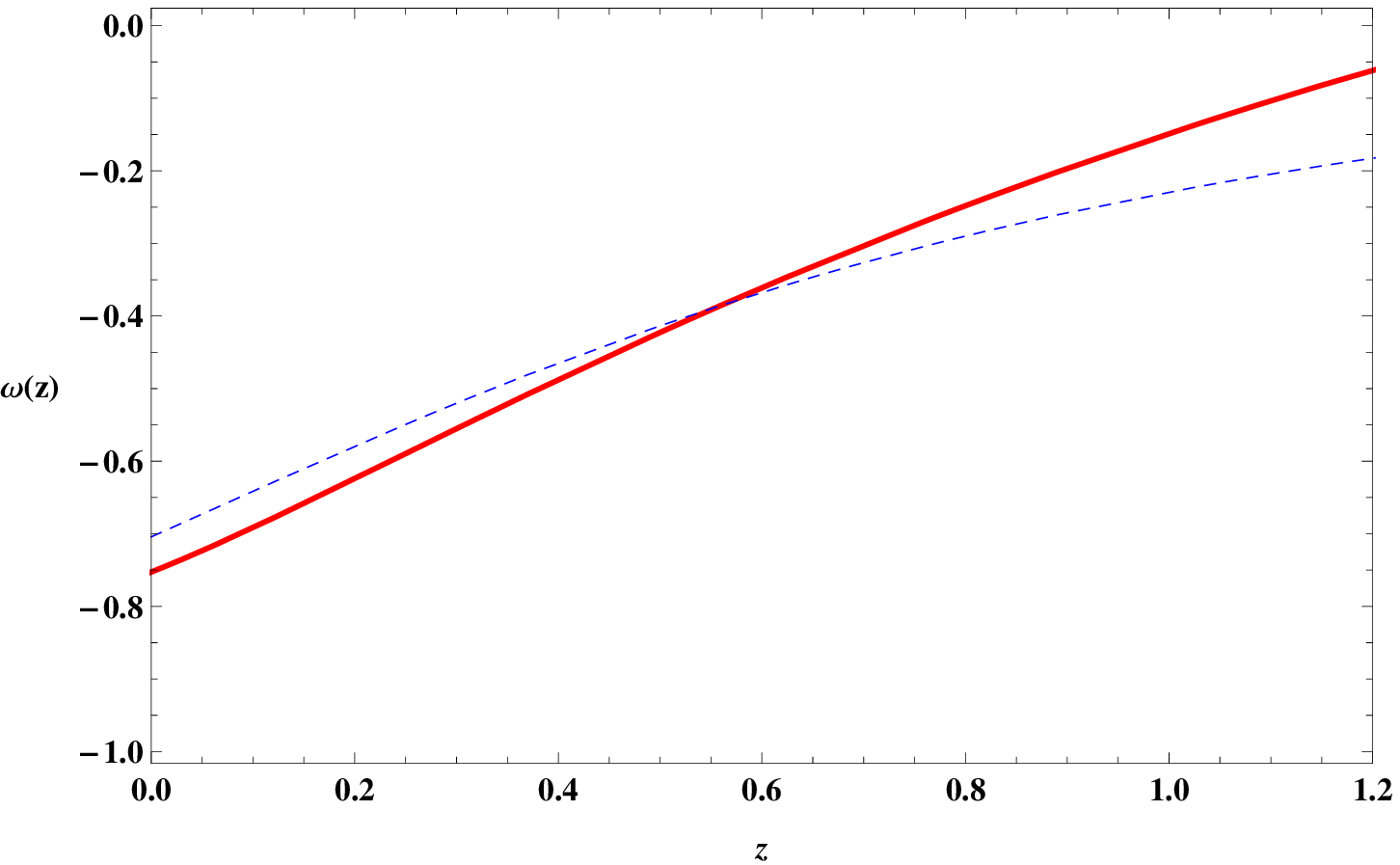}
\caption{\emph{Plot of the EoS. Again $q_1$ and $q_2$ are fixed to their best fit values of Tab. \ref{tabellaq1q2}. Solid line refers to the EoS of our model, while the dashed one to the $\Lambda$CDM paradigm, i.e. $\omega_{\Lambda CDM}=-[1+\frac{\Omega_{m,0}}{1-\Omega_{m,0}}(1+z)^3]^{-1}$. }}%
\label{omeg}}%
\end{figure*}

The fitting procedure is essentially described by employing the chi square definition $\chi$ as follows:
\begin{equation}\label{mod}
\chi^{2}_{super} =
\sum_{i}\frac{(\mu_{i}^{\mathrm{theor}}-\mu_{i}^{\mathrm{obs}})^{2}}
{\sigma_{i}^{2}}\,,
\end{equation}
where $\chi=\chi_{super}$, i.e. the chi square related to the supernova compilation. The apparent modulus, reported in Eq. (\ref{mod}), is defined as $\mu = 25 + 5 \log_{10}\left( \frac{d_L}{Mpc}\right)$ and the corresponding likelihood function $\mathcal{L}$ is expressed in terms of the chi square as $\mathcal{L}\propto \exp(-\chi_{super}^{2}/2)$. Our technique lies on minimizing the $\chi_{super}^{2}$, or equivalently to maximize the likelihood function. Our approach essentially differs from the $\chi^2$ proposed by \cite{dipi}, which makes use of nuisance parameters, since it considers only a standard maximization of the likelihood function. However, no significative cosmographic departures are expected in case of cosmographic fits, using different chi squares, as previously noticed in \cite{luongo10}.

\noindent For our numerical outcomes we used the available program \emph{Bayesian Analysis Toolkit}, represented by a $C++$  coded package. It employs toolkit for numerical analysis, which is essentially based on the Bayes’ theorem. The Monte Carlo procedure that we developed is built up through the relation:
\begin{equation}\label{poster}
\pi(\mathbf{p}|D) =\frac{\pi(D|\mathbf{p}) \pi_{0}(\mathbf{p})}
{\int \pi(D|\mathbf{p}) \pi_{0}(\mathbf{p}) d\mathbf{p}} \;\;
\end{equation}
where $\pi(\mathbf{p}|D)$ is the posterior distribution for the list of parameters $\mathbf{p}$ given data $D$.

\noindent To handle the corresponding contours, we assumed the interface given by the free program ROOT \cite{rootbat}.

\noindent The Markov Chain is produced by means of the Metropolis-Hastings algorithm \cite{metropol}. For the employed statistics, the union 2.1 survey provides 580 supernovae up to the redshift $z\in[0;1.414]$ which are used in our codes as initial data. Here, systematics is mostly negligible and does not significatively contribute to the error propagations. We derive the corresponding contour plots as reported in Figs. \eqref{ours}, \eqref{ours2}.

\noindent The parameter $q_2$ remains undetermined from theoretical limits, and so it becomes object of our numerical fits albeit its sign and a viable interval can be easily calibrated by reconstructing the behavior of the acceleration parameter in terms of current data, as proposed in \cite{pavon}.

\subsection{Numerical outcomes on cosmokinetic parameters}

\noindent For our purposes, we consider two fit procedures: the first has been employed by taking free all the cosmographic parameters, i.e. $H_0, q_0, j_0, q_1$ and $q_2$. The second has been obtained by fixing all the cosmographic parameters inside the best value intervals of $q_0$, $j_0$ and $q_1$, as indicated in Tab. \ref{tabellao}, which are compatible with the results of \cite{luongoprd2012}. In general, our cosmographic approach, by virtue of the Taylor treatments, even reduces degeneracy between matter and dark energy densities avoiding the circularity problem when one evaluates the luminosity distance.

\noindent In the first analysis, we check the goodness of our model. In the second analysis, we develop the fit between $H_0$ and $q_2$, since $q_1$ is univocally determined as $q_0$ and $j_0$ are known.

The numerical outcomes indicate that the values assumed by $H_0$ in both the fits are slightly larger than the Planck measurements \cite{planck}. The value of $q_0$ in the first fit is compatible with the $\Lambda$CDM prediction, whereas $j_0$ is larger than the one obtained in the concordance model, i.e. $j_0=1$. Indeed, we get $j_0>1$, with the $q_1$ magnitude perfectly inside the $\Lambda$CDM predictions, suggesting $|q_1|\sim 1$. The sign of $q_1$ is however different, while no conclusive results can be achieved concerning $j_1$ and $j_2$, that have been derived propagating the errors through the logarithmic rule. The values of $q_2$ lie on intervals centered around $q_2\sim1$. This prediction does not significatively change for both the two fits and seems to differ from the $\Lambda$CDM case. In other words, although the two fits seem to match with the standard paradigm at the level of $H_0$ and $q_0$, they provide slight differences: the sign of $q_1$, $q_2\sim 1\div2$ and $j_0>1$, which do not agree with the concordance model. Our two analyses are clearly not exhaustive to definitively conclude that dark energy evolves at the background level, but suggests a first indication of its possible evolution at higher redshift. In other words, the cosmographic model seems to be compatible with a \emph{weakly} evolving dark energy term, which reduces to the $\Lambda$CDM model in the limiting case $z\rightarrow0$. However, the use of combined cosmic data sets will clearly certify this fact, with additional numerical tests.

\begin{center}
\begin{table}[ht]
\begin{small}
\caption{{\small \emph{Summary of results obtained
fitting union 2.1 SNeIa data by means of Metropolis algorithm. Here,
$1\sigma$ errors are considered. To fix the Hubble parameter today a Gaussian prior over it has been considered, following \cite{exc}.}}}

\begin{tabular}{ccc}
\hline\hline

$\quad$$\qquad$$\quad$ & {\bf Union 2.1 Cosmographic Best Fit} &
$\quad$$\qquad$$\quad$\\

 \cline{2-2}

\end{tabular}

\begin{tabular}{ccc}\label{tabellaq1q2}

 $H_0$  &  & $p-value$  $$ \\

  $70.32\pm0.37$  &  & 0.687 \\[1ex]

  \hline

 $q_0$ $\qquad$ & $\qquad$ $q_1$ $\qquad$ & $\qquad$ $q_2$ \\[0.8ex]

 $-0.70\pm0.05$ $\qquad$ & $\qquad$ $1.54\pm0.34$ $\qquad$ & $\qquad$
$1.77\pm0.92$ \\[1ex]

 \hline

 $j_0$ $\qquad$ & $\qquad$ $j_1$ $\qquad$ & $\qquad$ $j_2$ \\[0.8ex]

 $1.82\pm0.44$ $\qquad$ & $\qquad$ $-0.78\pm3.13$ $\qquad$ & $\qquad$
$-2.04\pm5.99$ \\[1ex]

\hline\hline
\end{tabular}
\end{small}
\end{table}
\end{center}

\begin{center}
\begin{table}[ht]
\begin{small}
\caption{{\small \emph{Summary of results obtained for the considered
sample fitted exploiting the Metropolis algorithm. Here,
$1\sigma$ errors are employed. For $q_0$ and $j_0$ the corresponding errors are not reported because they are fixed values, compatible with previous analyses. Whereas for $q_1$ the errors vanish as a consequence of the standard logarithmic formula. To fix the Hubble parameter today a Gaussian prior over it has been considered, following \cite{exc}.}}}

\begin{tabular}{ccc}
\hline\hline

$\quad$$\qquad$$\quad$ & {\bf Union 2.1 Cosmographic Best Fit} &
$\quad$$\qquad$$\quad$\\

 \cline{2-2}

\end{tabular}

\begin{tabular}{ccc}\label{tabellaDUE}

 $H_0$  &  & $p-value$  $$ \\

  $70.50\pm0.24$  &  & 0.573 \\[1ex]

  \hline

 $q_0$ $\qquad$ & $\qquad$ $q_1$ $\qquad$ & $\qquad$ $q_2$ \\[0.8ex]

 $-0.65$ $\qquad$ & $\qquad$ $1.005$ $\qquad$ & $\qquad$
$1.42\pm0.40$ \\[1ex]

 \hline

 $j_0$ $\qquad$ & $\qquad$ $j_1$ $\qquad$ & $\qquad$ $j_2$ \\[0.8ex]

 $1.29$ $\qquad$ & $\qquad$ $0.01\pm 0.80$ $\qquad$ & $\qquad$
$-3.40\pm1.53$ \\[1ex]

\hline\hline
\end{tabular}
\end{small}
\end{table}
\end{center}


\section{Final outlooks and perspectives}

In this work, we revisited the degeneracy problem between matter and dark energy densities through the use of cosmography. Basically, cosmography can assess the degeneracy problem by simply considering the validity of the cosmological principle, under the hypothesis that the universe is spatially flat. Our procedure was to expand in Taylor series the acceleration $q$ and the jerk $j$ parameters, around current epoch, i.e. $a=1$. From this expansion, it has been easy to notice that matter density can be expressed in terms of the acceleration parameter today, whereas constraining $j$ at present time enables to describe the dark energy evolution at small redshift regimes. The consequence is that the whole content of the universe energy budget is framed in function of the cosmographic parameters. Since those terms are model-independent quantities, i.e. can be measured without imposing a dark energy model a priori, we inferred a cosmological dark energy model which does not include any \emph{ad hoc} assumptions. In other words, in lieu of considering matter and dark energy densities, one reduces the net number of free parameters by employing the cosmographic treatment within the Hubble and luminosity distance. In addition to current-time cosmographic terms $H_0$, $q_0$, $j_0$, we baptized further cosmographic quantities: $q_1, j_1$, $q_2, j_2$, related to the derivatives of $q$ and $j$. We formulated a cosmographic model which is dependent on those parameters and when $H_0$, $q_0$, $j_0$ are experimentally fixed, the model depends upon $q_2$ only. In fact, we showed that from a direct expansion of $d_L$ one can infer limits over the \emph{pure} cosmographic coefficients $H_0$, $q_0$, $j_0$. As a consequence, $q_1$ is determined through the formula $q_1\approx j-2q^2-q$, letting $q_2$ to freely vary as the unique unbounded parameter. By virtue of the Taylor treatments, the degeneracy between cosmographic coefficients less influences the whole analysis, contrary to the degeneracy between matter and dark energy densities which is jeopardized by the circularity problem when evaluating the luminosity distance. We also showed that to completely remove the degeneracy on the cosmographic couple of coefficients $q_1, q_2$, it is possible to compute $q_1$ in function of $\omega_1$, i.e. the first derivative with respect to the redshift of the universe EoS. We considered two cosmological fits with the union 2.1 compilation by means of Monte Carlo analyses, performed by the free available codes BAT and ROOT. In the first fit all the cosmographic coefficients were free to vary, while in the second only $q_2$ evolves. We found that the concordance model is confirmed within the 1$\sigma$ confidence level, although a slight evidence for an evolving dark energy contribution is not excluded by the analyses themselves. The numerical outcomes coming from the fits point out no conclusive indications for distinguishing the two cases. However, the strategy of splitting the measurement between $\Omega_{m,0}$ and $\omega$ is a robust hint on how to handle the dark energy contribution. In particular, we found that the cosmographic terms $H_0$ and $q_0$ are compatible with the standard cosmological model, the term $q_2$ is slightly smaller than the one predicted by the concordance model and $q_1$ provides a different sign. In addition, we got $j_0>1$ which differs from the $\Lambda$CDM case, confirming that the dark energy contribution would reduce to a constant only at $z\rightarrow0$, but evolving elsewhere. To this end, additional analyses will clearly reduce the error propagations by combining different data sets, in order to discriminate whether the model effectively provides an evolving dark energy term. In doing so, we expect to improve the accuracy in facing the degeneracy problem, showing how dark energy effectively changes in time or not. Future works, handling higher redshift terms in the Taylor expansions, may represent a key towards a more accurate cosmographic dark energy model, especially for what concerns its evolution at higher redshift domains.

\acknowledgements
O.L. is grateful to Manuel Scinta for his support in the numerical analyses and wishes to express his gratitude to prof. P. K. S. Dunsby for useful discussions. This work was partially supported by National Research Foundation (NRF).


\end{document}